\documentclass[reprint, superscriptaddress, amsmath, amssymb, aps, pra, a4paper]{revtex4-2}
\usepackage{graphicx}
\usepackage{dcolumn}
\usepackage{hyperref}
\usepackage{multirow}
\usepackage{comment}
\usepackage{xcolor}
\usepackage{amsmath}
\usepackage{mathrsfs}
\usepackage{tabularx}
\usepackage{natbib}
\usepackage{physics}
\usepackage{float}

\usepackage{soul}

\DeclareUnicodeCharacter{2009}{\,}

\begin{document}

\title{Passive polarization-encoded BB84 protocol using a heralded single-photon source}

\author{Anju Rani}
\email{galavanju1994@gmail.com}
\affiliation{Quantum Science and Technology Laboratory, Physical Research Laboratory, Ahmedabad, India 380009.}
\affiliation{Indian Institute of Technology, Gandhinagar, India 382355.}

\author{Vardaan Mongia}
\affiliation{Quantum Science and Technology Laboratory, Physical Research Laboratory, Ahmedabad, India 380009.}
\affiliation{Indian Institute of Technology, Gandhinagar, India 382355.}

\author{Parvatesh Parvatikar}
\affiliation{National Institute of Technology, Warangal, India 506004.}

\author{Rutuj Gharate}
\affiliation{Quantum Science and Technology Laboratory, Physical Research Laboratory, Ahmedabad, India 380009.}

\author{Tanya Sharma}
\affiliation{Quantum Science and Technology Laboratory, Physical Research Laboratory, Ahmedabad, India 380009.}
\affiliation{Indian Institute of Technology, Gandhinagar, India 382355.}

\author{Jayanth Ramakrishnan}
\affiliation{Quantum Science and Technology Laboratory, Physical Research Laboratory, Ahmedabad, India 380009.}

\author{Pooja Chandravanshi}
\affiliation{Quantum Science and Technology Laboratory, Physical Research Laboratory, Ahmedabad, India 380009.}

\author{R. P. Singh}
\email{rpsingh@prl.res.in}
\affiliation{Quantum Science and Technology Laboratory, Physical Research Laboratory, Ahmedabad, India 380009.}

\date{\today}

\begin{abstract}
The BB84 quantum key distribution protocol set the foundation for achieving secure quantum communication. Since its inception, significant advancements have aimed to overcome experimental challenges and enhance security. In this paper, we report the implementation of a passive polarization-encoded BB84 protocol using a heralded single-photon source. By passively and randomly encoding polarization states with beam splitters and half-wave plates, the setup avoids active modulation, simplifying design and enhancing security against side-channel attacks. The heralded single-photon source ensures a low probability of multi-photon emissions, eliminating the need for decoy states and mitigating photon number splitting vulnerabilities. The quality of the single-photon source is certified by measuring the second-order correlation function at zero delay, $g^{2}(0)=0.0408\pm0.0008$, confirming a very low probability of multi-photon events. Compared to conventional BB84 or BBM92 protocols, our protocol provides optimized resource trade-offs, with fewer detectors (compared to BBM92) and no reliance on external quantum random number generators (compared to typical BB84) to drive Alice's encoding scheme. Our implementation achieved a quantum bit error rate of 7\% and a secure key rate of 5 kbps. These results underscore the practical, secure, and resource-efficient framework our protocol offers for scalable quantum communication technologies.
\end{abstract}

\maketitle

\section{\label{sec:Introduction}Introduction}
Quantum key distribution (QKD) is one of the most ubiquitous applications of quantum information, which exploits the fundamental principles of quantum mechanics to create a shared cryptographic key between two communicating parties. Unlike classical cryptographic methods, QKD provides information-theoretic security, relying entirely on quantum laws rather than mathematical complexity, ensuring robustness against adversaries with unlimited computational power.  
\par The BB84 protocol\,\cite{Bennett1984}, the first QKD protocol, encodes key information in the polarization states of single photons. Since its inception, various protocols, such as E91\,\cite{E91}, BBM92\,\cite{BBM92}, DPSK\,\cite{DPSK}, and COW\,\cite{COW}, have been introduced, each promising theoretical unconditional security\,\cite{Lo1999,Shor2000,GLLP}. However, practical implementations face challenges, including experimental imperfections that introduce vulnerabilities against security\,\cite{Huttner1994,Brassard2000,Makarov2005}. For instance, implementing a typical BB84 protocol requires a single-photon source capable of producing one photon per pulse deterministically. However, generating a true single-photon is experimentally challenging\,\cite{2023NatRP...5..326C}. As a result, weak coherent pulses following Poissonian statistics are often used, which introduces a non-zero probability of multi-photon emissions, making the protocol vulnerable to photon number splitting (PNS) attacks\,\cite{Brassard2000}. To address these vulnerabilities, innovative protocols such as decoy state protocols\,\cite{lo2005decoy} and entanglement-based (EB) protocols\,\cite{E91,BBM92} have been introduced. However, these protocols often increase implementation complexity. Decoy state BB84 protocols often use multiple lasers to generate signal and decoy states, increasing vulnerability to side-channel attacks. Moreover, the use of intensity modulators to adjust the mean photon number increases the setup's size and complexity. While EB protocols offer higher security, they suffer from low key generation rates due to entanglement verification and face challenges in distributing entanglement over long distances with high fidelity.
\par There is a great interest in developing fully passive QKD transmitters\,\cite{Zapatero_2023,PhysRevA.82.052325,PhysRevApplied.14.024036}– i.e. devices that do not need to actively change the encoding state or its mean photon number, but instead rely on passive components such as beam splitters and fixed-path interferometers to determine the output states randomly. For example, in the standard BB84 decoy state protocol, it is necessary to encode the quantum states in a chosen observable—such as phase, polarization, or time-bin—and implement decoy states, which involve random variations in pulse intensity, to maintain security against the non-zero probability of transmitting multi-photon states. The advantages of passive systems include (i) simpler construction, longer performance stability, and reduced maintenance; (ii) enhanced resistance to Trojan-horse attacks\,\cite{PhysRevX.5.031030} that exploit vulnerabilities in active device settings; and (iii) minimal operational fingerprints at high emission rates (GHz), which are challenging to suppress in active systems and could be exploited by attackers\,\cite{10.1063/5.0021468}. These factors make passive QKD transmitters a compelling alternative for secure quantum communication.
\par In this paper, we have reported the implementation of passive polarization-encoded BB84 protocol using a heralded single-photon source. The implementation involves passively and randomly encoding of the polarization of transmitted pulses in the QKD transmitter (Alice). In addition, the photon source used is a heralded single-photon source, where, the time of detection of the heralding photon is used to identify the transmitted polarization state from knowing the time of its detection by the QKD receiver (Bob). To evaluate the quality of the single-photon source, we measured the second-order correlation function at zero delay, $g^{2}(0)$\,\cite{agarwal2012quantum,Lal2019SinglePS}. The measured single-photon-ness i.e. the lower value of $g^{2}(0)$ confirms that the heralded photon states have a very low probability of being multi-photon, ensuring the security of the implemented protocol, thereby removing the need to create decoy states.
\par Our approach offers several advantages over conventional BB84 or decoy state implementations. The use of the randomly-emitted heralding signal photon to determine the polarization of the detected photon (the time-correlated idler photon, after passing through the passive encoding scheme) eliminates the need for side-channel and hence is secure against the side-channel attacks\,\cite{Ayan}. Using a heralded single-photon source with a lower value of $g^{2}(0)$ mitigates vulnerabilities to PNS attacks, while the absence of decoy states simplifies the setup. Additionally, our protocol requires fewer resources than BBM92, reducing the number of detectors at the receiver's end from eight to five\,\cite{Mishra_2022}. The protocol also avoids reliance on external quantum random number generators (QRNG)\,\cite{Mannalatha_2023,Gehring2021HomodynebasedQR}, instead using a beam splitter for random state selection.
\par This paper is organized as follows: Section\,\ref{sec:Theory} provides the theoretical background and outlines the security parameters of the protocol. Section\,\ref{sec:RnD} details the experimental implementation of the passive polarization-encoded BB84 protocol using a heralded single-photon source and presents the experimental results. Finally, Section\,\ref{sec:Conclusion} concludes with key remarks.

\section{\label{sec:Theory}Background}
This section describes the implementation of the passive polarization-encoded BB84 protocol using a heralded single-photon source. At Alice's end, single-photon pairs are generated via the spontaneous parametric down-conversion (SPDC) process\,\cite{Steinlechner_2012}, where a pump photon is annihilated to produce signal and idler photons, conserving energy and momentum. The signal photon is used for heralding, while the idler photon is used to encode the polarization states $|H\rangle$, $|V\rangle$, $|D\rangle$, and $|A\rangle$. These states are transmitted to Bob through a free-space channel, where the measurements are performed using single-photon detectors (SPD). The process is detailed in the following steps.

\subsection{Random Selection of the States}
In typical BB84 and decoy state protocols, a QRNG is used to generate random numbers for state selection. In the heralded BB84 protocol, this randomness is embedded into the setup itself. 
\begin{figure}[h!]
		\centering		
        \includegraphics[width=0.95\linewidth]{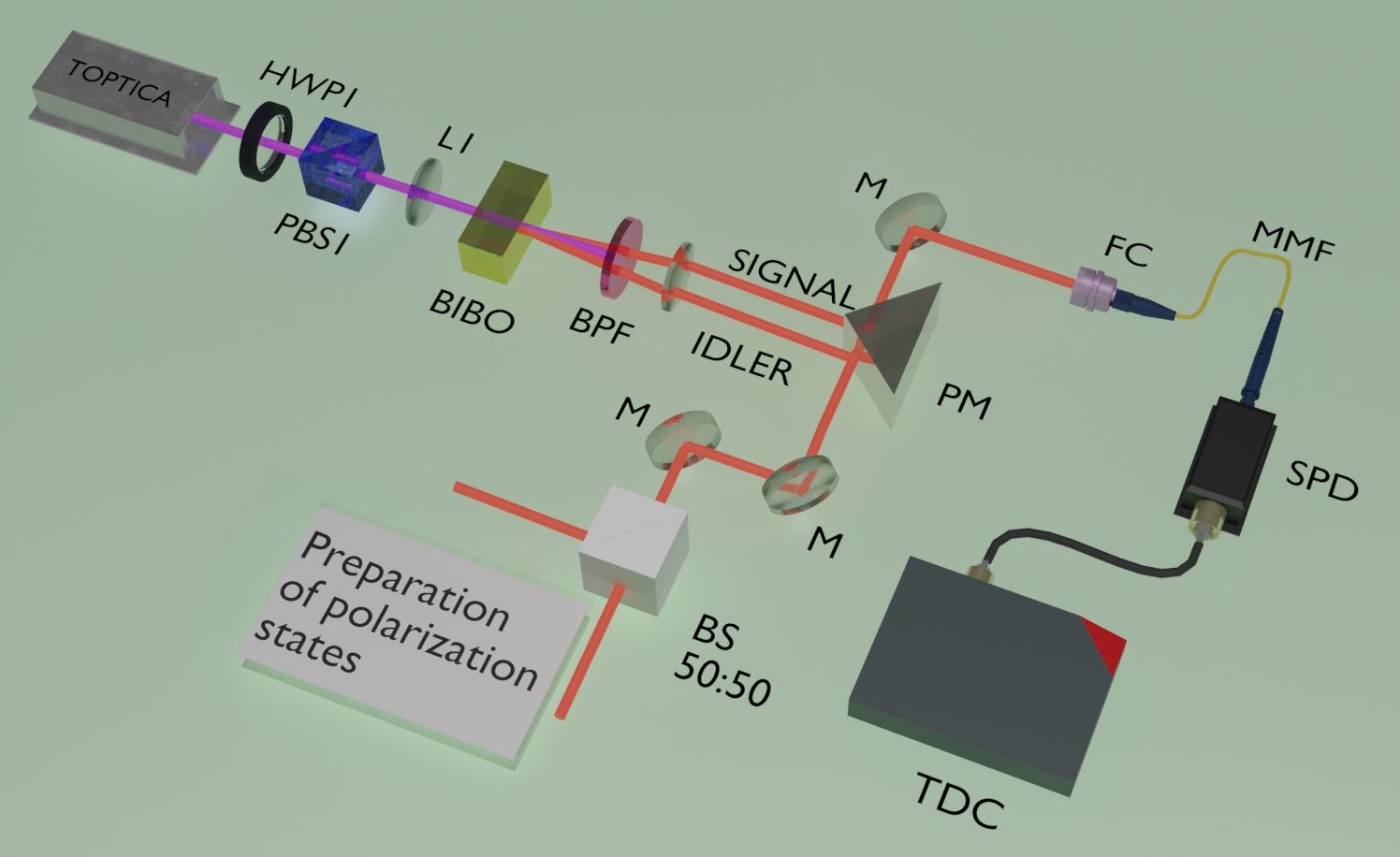}
		\caption{Schematic for generating heralded single photons using SPDC process. The signal photon is heralded, whereas, the idler photon is used to encode the four polarization states; HWP: Half-wave plate; PBS: polarizing beam splitter; L: Lens; BIBO: Bismuth Borate nonlinear crystal; BPF: Band-pass filter; PM: Prism mirror; FC: Fiber coupler; MMF: Multi-mode fiber; SPD: Single-photon detector; TDC: Time-to-digital converter; BS: Beam splitter.}
 \label{fig:SPDCCC}
\end{figure}
\par The experimental setup for generating heralded single-photons using the SPDC process is shown in Fig.\,\ref{fig:SPDCCC}. Alice uses a 405 nm continuous-wave diode laser (TOPTICA) with 6 mW output power. A combination of a half-wave plate (HWP1) and a polarizing beam splitter (PBS1) is used to regulate the pump power delivered to the type-I Bismuth Borate (BIBO) crystal, which generates single photon pairs using the SPDC process. A lens (L1) of focal length 50 cm is used to focus the pump beam onto the crystal, enhancing SPDC efficiency. The vertically polarized signal and idler photons are emitted from the crystal at 810 nm, which are separated by using a prism mirror (PM). A band-pass filter (BPF) for 810 nm wavelength having a bandwidth of 10 nm is used to block the 405 nm pump beam. The signal photon, used for heralding, is coupled to a fiber coupler (FC) via a multi-mode fiber (MMF) and detected by a single-photon detector (Excelitas, SPCM-AQRH-14-FC). time stamps are recorded using a time-to-digital converter (ID800 TDC, ID Quantique). The idler photon is used to prepare the four polarization states. The BSs placed in the idler arm do the job of randomly selecting the photons sent to Bob. 
\begin{figure}[h!]
		\centering		
        \includegraphics[width=0.95\linewidth]{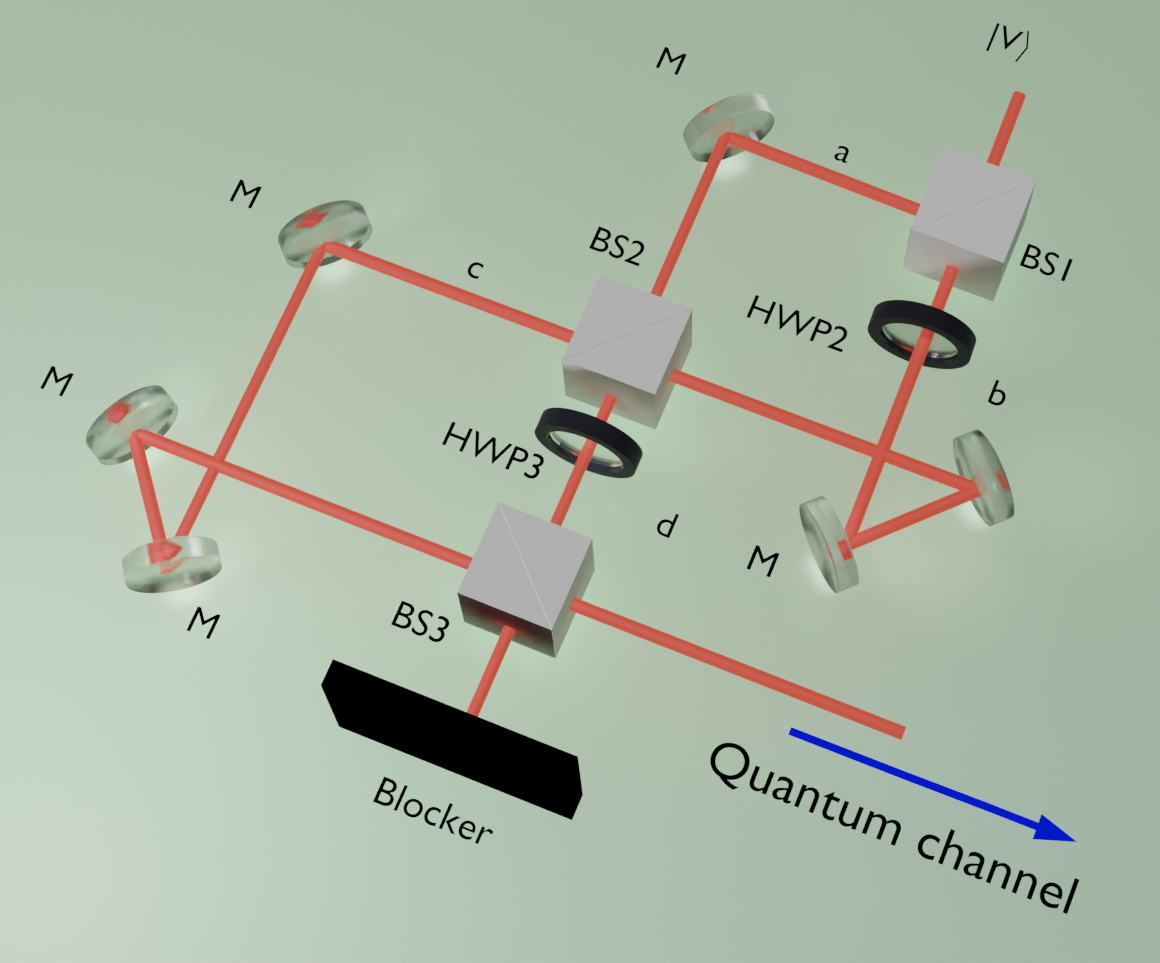}
		\caption{Passive polarization state preparation for BB84 protocol using one of the heralded single photons (idler). Two cascaded Mach-Zehnder interferometers are set up, with HWP in them, to prepare $|H\rangle$, $|V\rangle$, $|D\rangle$, and $|A\rangle$ polarization states; HWP: Half-wave plate; BS: Beam splitter; M: Mirrors.}
  \label{fig:encodingg}
    \end{figure} 
\par The polarization states $|H\rangle$, $|V\rangle$, $|D\rangle$, and $|A\rangle$ are prepared using two cascaded Mach-Zehnder interferometers with HWPs in them, as shown in Fig.\,\ref{fig:encodingg}. A single photon can take one of four paths having different lengths: ac, bc, ad, or bd, randomly determined by beam splitters (BS1 and BS2). The BSs are ensured to be as close to 50:50 as possible, ensuring equal probability for each state. The initial polarization at BS1 is $|V\rangle$. The HWP2 and HWP3 are placed at $45^{\mathrm{o}}$ and $22.5^{\mathrm{o}}$ to convert the initial polarization into $|H\rangle$ and $|D\rangle$, respectively. The photon following the path 'ac' or 'bc' is $|V\rangle$ or $|H\rangle$ polarized. The photon following the paths 'ad' or 'bd' is $|A\rangle$ or $|D\rangle$ polarized. The photons encoded in different polarizations, arriving at BS3, are transmitted to Bob with a 50\% transmission rate. The single-photon interference at BS3 is ruled out as the photons encoded in different polarizations, travel through different arms of unequal lengths, making the paths non-identical.
\subsection{Transmission and Detection of States}
The quantum state is transmitted to Bob through a quantum channel. One can use free space or optical fiber as a channel for sending qubits in polarization degrees of freedom. Free space offers an advantage as the polarization drifts in free space are much less compared to fiber. Exploring free space is essential for terrestrial applications and satellite communication. 
\par
\begin{figure}[h!]
		\centering		
        \includegraphics[width=0.95\linewidth]{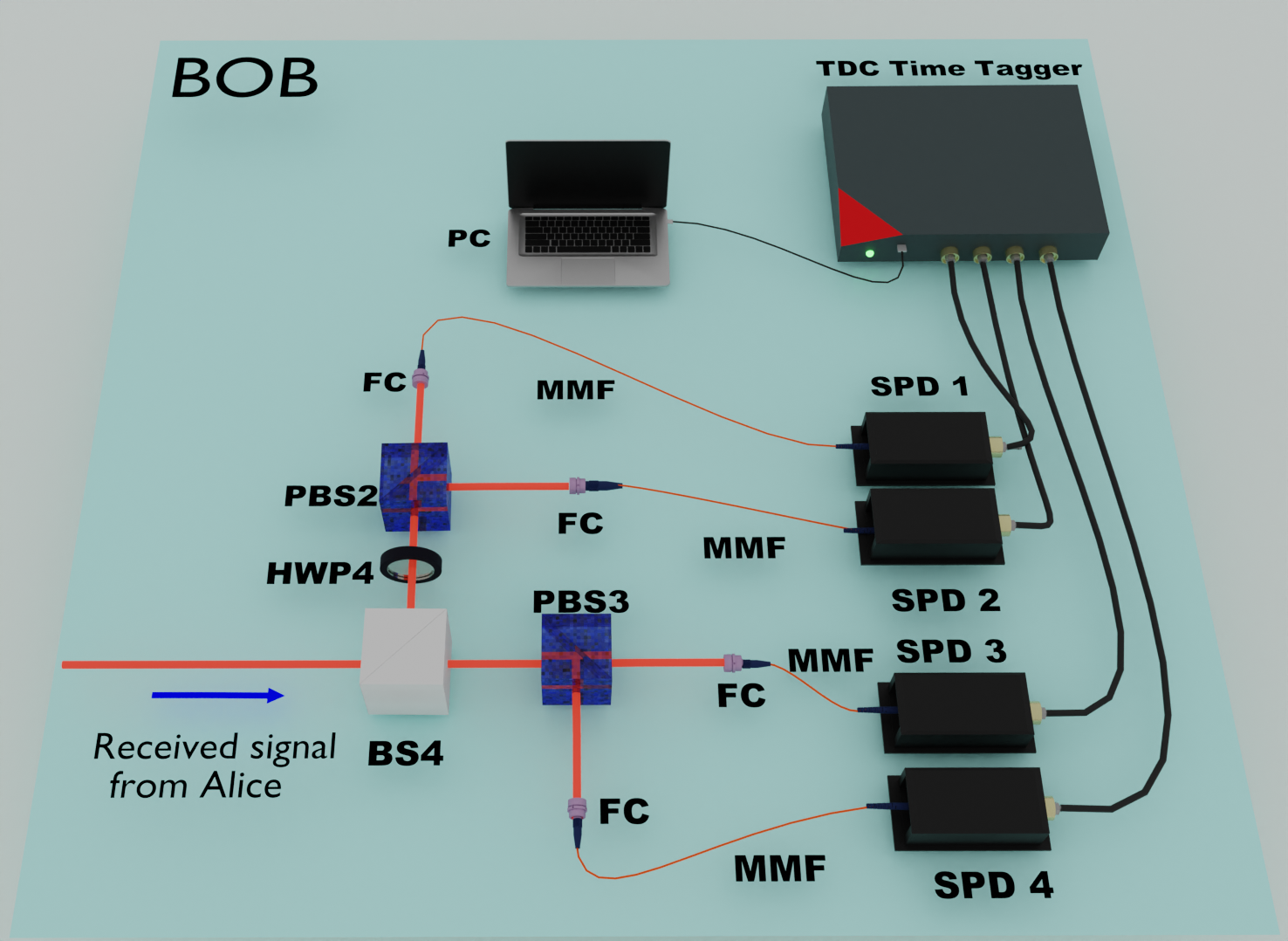}
		\caption{Detection of the quantum states at receiver's end; HWP: Half-wave plate; PBS: polarizing beam splitter; FC: Fiber coupler; SPD: Single-photon detector; TDC: Time-to-digital converter; BS: Beam splitter.}
  \label{fig:bobb}
    \end{figure}

 The photons are measured by projecting their polarization state onto the four basis states by employing a combination of HWPs and PBSs followed by SPDs. Bob's detection setup includes a 50:50 beam splitter (BS4), which acts as a basis selector that randomly selects the basis for the projection measurement, as shown in Fig.\,\ref{fig:bobb}.
The state measurement in the $\{H, V\}$ basis is performed using PBS3, while in $\{D, A\}$ basis is performed using a combination of HWP4 and PBS2. The photons are coupled to MMF and detected by SPDs. The detected photon counts and their arrival times are recorded using TDC. 

\subsection{Post-processing}
Once the key establishment process is done, Alice and Bob further process their data to extract the secure key. They perform sifting to get the correlated bit string. The sifting process requires that the photon arrival time at the receiver's end and the time when they are launched from the transmitter must be known. Alice and Bob are inherently synchronized due to the SPDC process and thus the time stamps information can be correctly extracted by both Alice and Bob. Since Alice has encoded her state in polarization states following different optical paths, she precisely knows the delays associated with each polarization state, as summarized in Table\,\ref{table:Pol_delay}. Moreover, the absolute delay of the heralded signal photon from the crystal to the detector is known by Alice, which is $3.07$ ns. To perform sifting, Alice and Bob follow the steps mentioned here.
\begin{table}[h]
    \centering
    \begin{tabular}{|c|ccc|}
        \hline
        Path & \multicolumn{1}{c|}{Absolute Time Delay (ns)} & \multicolumn{1}{c|}{Output States}
        \\ \hline
        ac & \multicolumn{1}{c|}{$11.84$}& \multicolumn{1}{c|}{$|V\rangle$}
        \\ \hline
        ad & \multicolumn{1}{c|}{$8.8$}& \multicolumn{1}{c|}{$|A\rangle$} 
        \\ \hline
        bc & \multicolumn{1}{c|}{$13.57$} & \multicolumn{1}{c|}{$|H\rangle$}
        \\ \hline
        bd & \multicolumn{1}{c|}{$10.52$}& \multicolumn{1}{c|}{$|D\rangle$}
        \\ \hline
    \end{tabular}
    \caption{The table contains time delays for different output polarization states at Alice's end, and this delay information is limited to Alice only.}
    \label{table:Pol_delay}
\end{table} 
\begin{itemize}
    \item Bob sends his time stamps and corresponding basis information to Alice. Alice compares the time stamps obtained for the heralded signal with Bob’s time stamps.
    \item The relation between the recorded time stamps of Alice\,($t_{A}$) and Bob\,($t_{B}$)  follow the relation 
    \begin{equation}
        t_{B} = t_{A} + \Delta + \delta_{ch}
    \end{equation}
    where $\Delta$ is the delay (in ns units) in the path lengths in the state preparation setup which is given in Table\,\ref{table:Pol_delay} for each polarization, and $\delta_{ch}$ is the delay due to the quantum channel which is fixed for all polarizations.  
    \item If the difference in the recorded time stamps matches the time delay for any of the four polarization states, then the time stamps correspond to the state sent by Alice.
    \item Alice removes all the detections for which the time stamp difference is not one of the four time delays mentioned in Table \ref{table:Pol_delay}. Also, the detections corresponding to the wrong basis measurements are discarded.
    \item Alice and Bob are left with correlated key elements (sifted key).  
\end{itemize}
\par The quantum bit error rate (QBER) is then evaluated using a small portion of the sifted key. The estimated QBER is used to evaluate the mutual information shared between Alice and Bob and also to further assess the amount of information leaked to a potential Eve. We utilized low-density parity check (LDPC) codes with a code rate of 0.5 for error correction in our implementation. The procedure followed aligns with the steps detailed in \cite{ldpc-QKD} for integrating LDPC codes into QKD systems. Privacy amplification is performed using Toeplitz hashing\,\cite{PhysRevA.87.062327, 10.1007/978-3-540-30576-7_22}.

\subsection{Security of the Protocol: $g^{(2)}(0)$ Correlation}
\begin{figure}[h!]
		\centering		
        \includegraphics[width=0.95\linewidth]{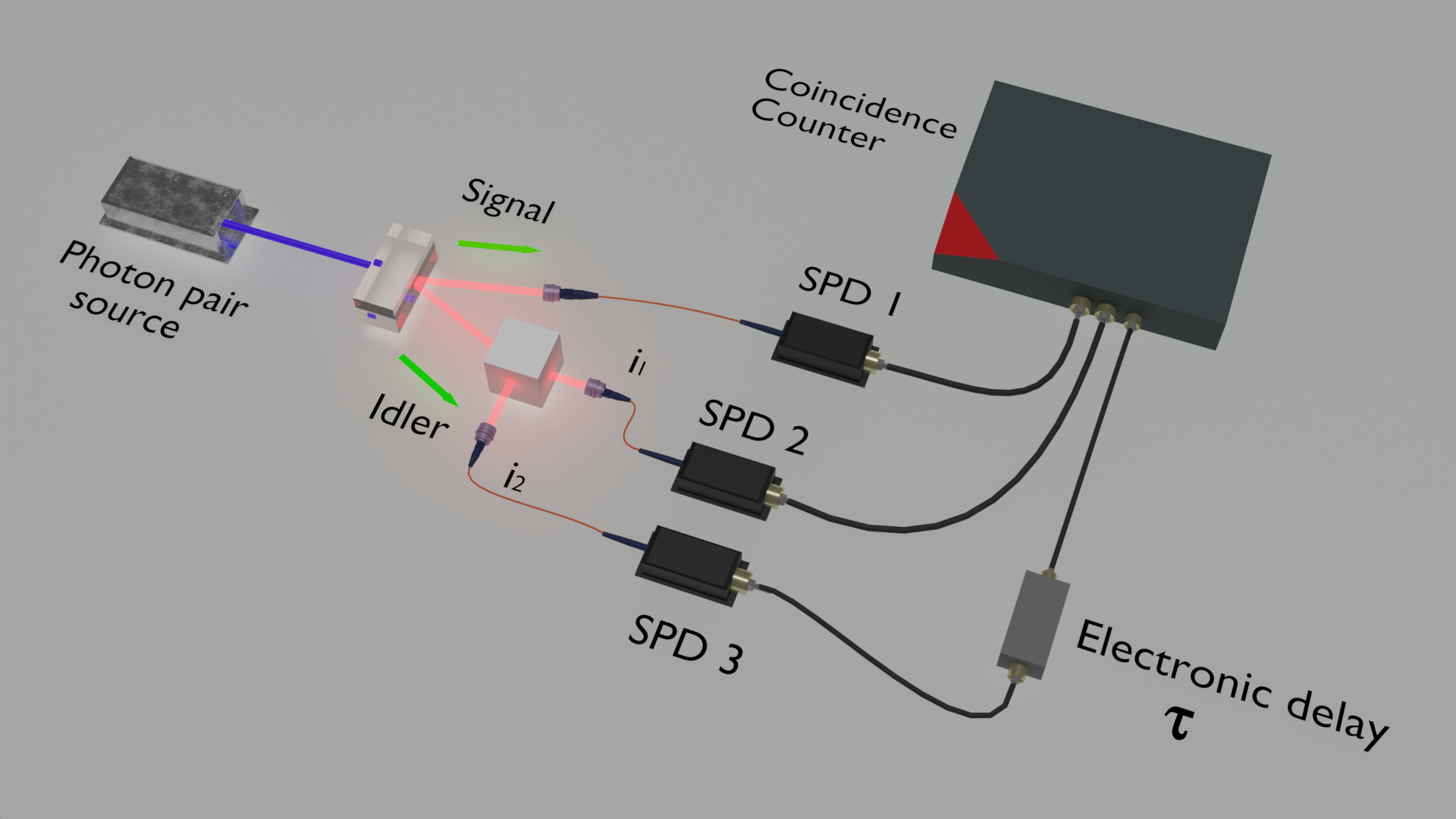}
	\caption{An illustration of the HBT experiment to determine the second-order correlation for a heralded single-photon source.}
    \label{fig:g22}
    \end{figure}

In quantum optics, the measurement of the second-order correlation $g^{(2)}(\tau)$ plays an important role, particularly in the observation of a purely quantum phenomenon called 'anti-bunching'\,\cite{gerry_knight_2004}. The normalized second-order correlation function\,\cite{nijil}, for a fixed position, is given as
\begin{equation}\label{eq:ng20}
    g^{(2)}(\tau) = \frac{\langle{{n}_{1}(t)}{{n}_{2}(t+\tau)}\rangle}{\langle{{n}_{1}(t)}\rangle\langle{{n}_{2}(t+\tau)}\rangle}
\end{equation}
where $n_{x}(t)$ and $n_{x}(t+\tau)$ are the number of counts detected at time '$t$' and '$t+\tau$' in the respective detectors. This measurement is commonly carried out using the Hanbury Brown and Twiss (HBT) experiment\,\cite{agarwal2012quantum}. In the case of an ideal single-photon source, where photons are emitted sequentially, each photon faces the option of being transmitted through the BS or being reflected. When the two paths have the same length, the probability of getting the clicks in two detectors simultaneously is zero, i.e. $g^{(2)}(0) =0$. This confirms the true single-photon nature of the source and ensures the security of the QKD experiments. For heralded single-photon sources produced by SPDC, the correlation of the photons in the HBT experiment is performed between the idler $(i)$ and the conditioned detection of the signal $(s)$, as depicted in Fig.\,\ref{fig:g22}.
In the perfect case of a single-photon source, where there is no delay between the detections of photons in arms $s$, $i_1$, and $i_2$, the three-fold detection probability among these arms becomes zero. This probability, represented as $P_{s,i_1,i_2}$, when normalized with respect to the corresponding two-fold coincidences, provides the second-order correlation for zero delay\,\cite{nijil}.
\begin{equation}\label{eq:g2}
    g^{(2)}(0) = \frac{P_{s,i_1,i_2}}{P_{s,i_1}P_{s,i_2}}
\end{equation}
where $P_{s,i_1}$ and $P_{s,i_2}$ are the probabilities of two-fold coincidence between $s-i_1$, and $s-i_2$ pairs. The expression for the conditioned probability of coincidence detection is given by $P_{i, j} = \frac{R_{i,j}}{R_{i}}$, where $R_{i,j}$ represents the coincidence rates in the respective arms, and $R_i$ represents the count rate of the heralding arm. By substituting various probabilities into Eq.\,\ref{eq:g2}, the expression for $g^{(2)}(0)$ in terms of experimentally measured rates can be represented as follows\,\cite{nijil}:
\begin{equation}\label{eq:g20}
    g^{(2)}(0) = \frac{R_{s,i_1,i_2} R_{s}}{R_{s,i_1}R_{s,i_2}}
\end{equation}
To calculate $g^{(2)}(0)$, one can directly measure the three-fold coincidences between the heralding signal mode and the two idler modes.
\par The experimental setup for studying the correlation of heralded single photons from the SPDC process is depicted in Fig.\,\ref{fig:g22}. As detailed in the previous section, degenerate signal–idler photon pairs are generated through the SPDC process. The signal photons are coupled into a MMF, while the idler photons are split equally using a 50:50 BS, with each output coupled into separate MMFs. The signal and idler photons from the two BS ports are directed to SPDs. Fiber coupling is optimized using fiber collimators to maximize coincidences between the signal and idler arms. Coincident photon detection is performed using TDC. The detector positions, fiber lengths, and BNC connectors from the SPDs to the TDC are carefully adjusted to ensure zero relative delays between the three detectors. This alignment is verified by achieving maximum two-fold coincidences between $s-i_1$, and $s-i_2$.
\section{\label{sec:RnD}Results and Discussion}
\begin{figure}[h!]
		\centering		
        \includegraphics[width=01\linewidth]{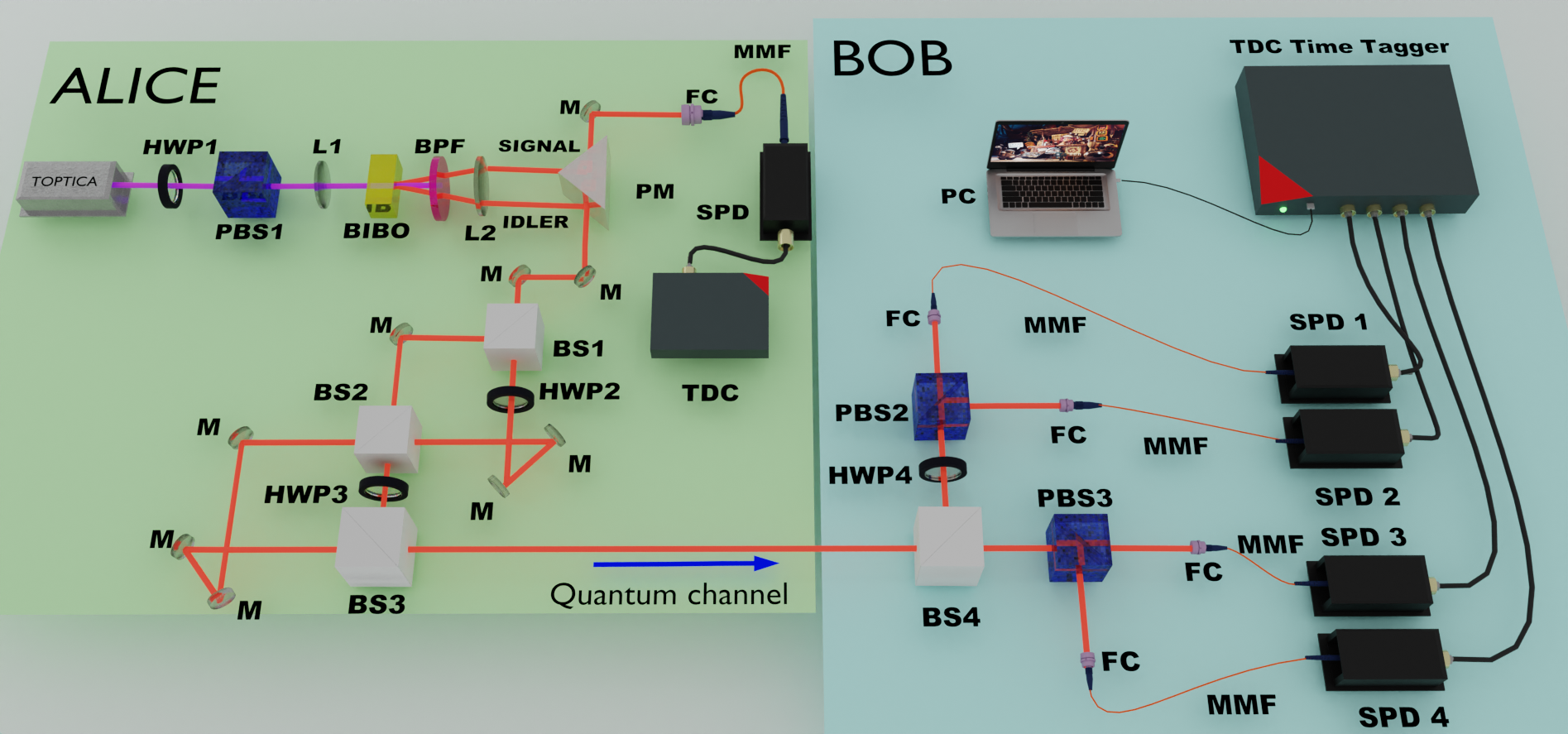}
		\caption{Experimental setup for passive polarization-encoded BB84 protocol using heralded single-photon source; HWP: Half-wave plate; PBS: polarizing beam splitter; L: Lens; BIBO: Bismuth Borate nonlinear crystal; BPF: Band-pass filter; PM: Prism mirror; FC: Fiber coupler; MMF: Multi-mode fiber; SPD:  Single-photon detector; TDC: Time-to-digital converter; BS: Beam splitter; M: Mirrors.}
  \label{fig:fulll}
    \end{figure}

We successfully implemented the passive polarization-encoded BB84 protocol using a heralded single-photon source over a free-space channel in our laboratory. The comprehensive experimental schematic for the protocol is illustrated in Fig.\,\ref{fig:fulll}, with detailed descriptions provided in the preceding section. 
\par A Type-I BIBO crystal of length 5 mm was used. The pump power before the nonlinear crystal was 6 mW. Temperature stabilization was not required for the crystal. During the experiment, we assumed that the BSs employed were nearly 50:50. This ensured that the basis measurements were unbiased, making both basis equally probable, which is critical for the integrity and fairness of the QKD process. The detector efficiencies were assumed to be similar, with a value of 65\%, specified in the manufacturer's data sheets. The detector's dark count rate was approximately 100 counts per second and was consistent for all detectors.
\par The channel length between Alice and Bob was 75 cm, and the measured channel transmittance in the lab for the free-space setup was 98\%. Additionally, the coupling efficiency of the MMF was determined to be 85\%. The experiment was conducted for an acquisition time of 1 second and a detection window of 1.5 ns. The obtained sifted key rate was 14 kbps, with a QBER of 7\%. After applying error correction and privacy amplification, the secure key rate\,\cite{PhysRevLett.100.090501} obtained was 5 kbps per raw sifted bit.
\par
\begin{figure}[h!]
		\centering		
        \includegraphics[width=0.85\linewidth]{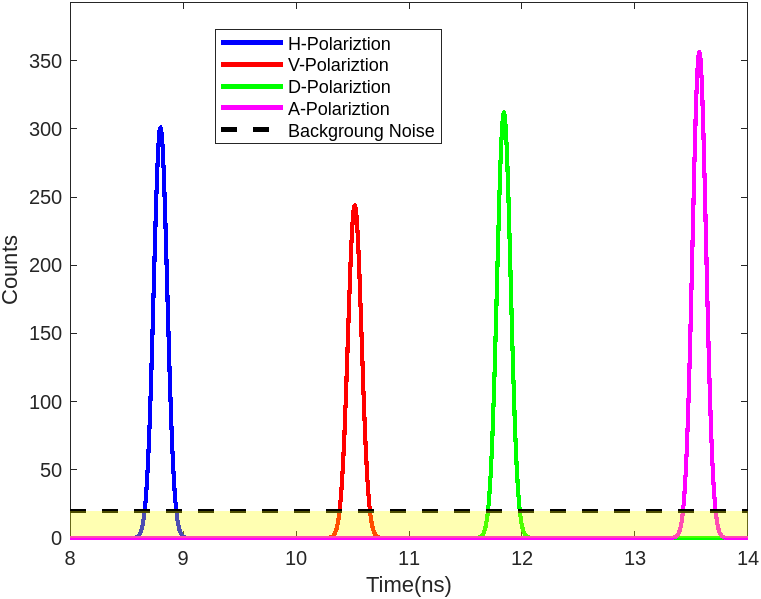}
		\caption{Output of the four independent correlated detections performed for Alice and Bob basis. The peaks indicate the correlated counts for the respective polarizations. The yellow zone indicates the background counts that represent the unwanted detections due to stray light or uncorrelated signal and idler photons.}
  \label{fig:anju1}
    \end{figure}

 We plotted the correlated counts obtained from the independent detections performed by Alice and Bob. The correlation is established between the heralded photon measured by Alice and the polarization-encoded photons received by Bob at different time delays. The time delay for the heralded photon is fixed, which is 3.07 ns. The photons at Bob's end arrive at different delays. The correlated counts for various measured polarizations are plotted in Fig.\,\ref{fig:anju1}. From the Figure, we can see that the correlated counts are less for V-polarization. This could be due to the experimental imperfections or poor coupling of the polarized photons to the fiber.
\par
\begin{figure}[h!]
		\centering		
        \includegraphics[width=1\linewidth]{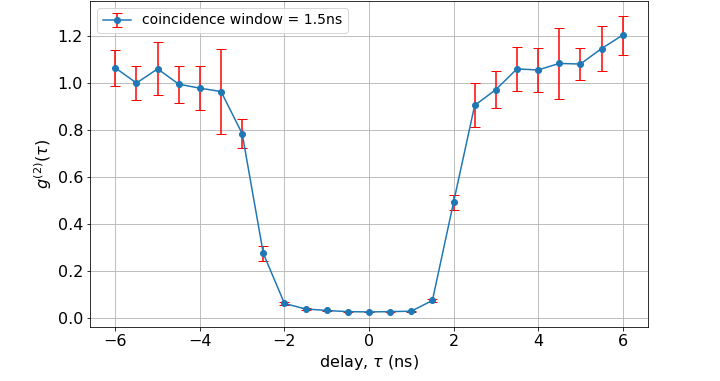}
		\caption{The variation of second-order correlation function $g^{(2)}(\tau)$ with delay, $\tau$. The error bar corresponds to the standard deviation obtained from the measured count rate uncertainties.}
  \label{fig:finalg2curve}
    \end{figure}

 Further, we performed the calculations for the second-order correlation function at Alice's end using Eq.\,\ref{eq:g20} as a function of the time delay between photons reaching the BS, shown in Fig.\,\ref{fig:g22}. A delay generator introduces electronic delays in 0.5 ns steps between the two output ports of the BS ($i_1$ and $i_2$) in the idler arm. Coincidence measurements between detectors SPD2 and SPD3 in these arms are heralded by photon detection in the signal arm. Anti-bunching (for $\tau = 0$) is observed by varying the temporal delay between $i_1$ and $i_2$, as shown in Fig.\,\ref{fig:finalg2curve}.
\par The measured value of  $g^{(2)}(0)$ for MMF is $g^{2}(0)=0.0408\pm0.0008$. While the ideal value of $g^{(2)}(0)$ for a true single-photon source is zero, non-zero values observed in heralded detection can be attributed to accidental coincidences among the three detectors. These coincidences may result from the simultaneous generation of multiple photon pairs, dark counts, or background fluorescence. Moreover, to maintain the non-classical behavior of the source, the pump power is kept in the lower regime (pump power=6 mW). However, the remarkably low value of $g^{(2)}(0)$ obtained from the experiment confirms an exceptionally minimal probability of multi-photon events, thereby ensuring the protocol's robustness against PNS attacks.
\par The final secure key rate is relatively low, primarily due to losses in the experimental setup. A significant factor is that 50\% of the transmitted signal is lost at BS3 (50:50), as shown in Fig.\,\ref{fig:fulll}. Additionally, the yield of the SPDC process itself is low, reducing the coincidence counts. The pump power is deliberately kept low to maintain a low heralded single-photon flux, which helps avoid multi-photon events. Future enhancements may include employing brighter entangled photon sources\,\cite{jabir2017robusthighbrightnessdegenerate} to boost the key rate. Ongoing efforts aim to further reduce the QBER of the protocol. Future research will also focus on evaluating the system's resistance to potential attacks and performing a more thorough security analysis to ensure the protocol's robustness against experimental impairments.
\par Despite the aforementioned limitations, this experiment effectively demonstrates, in principle, that our protocol provides security against side-channel attacks and PNS attacks and hence offers greater security compared to conventional BB84 or decoy state protocols. Furthermore, our implementation is both less resource-intensive and easier to realize.
\section{\label{sec:Conclusion}Conclusion}
In conclusion, we have successfully implemented a passive polarization-encoded BB84 QKD protocol using a heralded single-photon source over a free-space channel. The experimental setup demonstrated the feasibility of secure key distribution without the need for active state modulation, reducing system complexity and enhancing resistance to side-channel attacks. The measured QBER of 7\%, the sifted key rate of 14 kbps, and the secure key rate of 5 kbps highlight the protocol's practical viability. The use of a heralded single-photon source further ensured the low probability of multi-photon emissions, which is certified by the measured value of $g^{2}(0)=0.0408\pm0.0008$, making the system secure against PNS attacks. Overall, our work provides a promising framework for scalable, secure quantum communication systems with reduced resource requirements and improved operational stability. 

\section*{\label{sec:Acknowledgments}Acknowledgments}
The authors acknowledge the financial support from DST through the QuST program.

\section*{\label{sec:Disclosure}Disclosures}
The authors declare no conflicts of interest related to this article.


\bibliography{reference}

\end{document}